# Band Structure and Transport Studies on PdAl$_2$Cl$_8$-Intercalated Graphite

R.T.F. van SCHAIJK[a], A. de VISSER[a], E. McRAE[b], B. SUNDQVIST[c], T. WAGBERG[c], R. VANGELISTI[b]

[a] Van der Waals-Zeeman Institute, University of Amsterdam, Valckenierstraat 65, 1018 XE Amsterdam, The Netherlands;
[b] Laboratoire de Chimie du Solide Minéral, Université Henri Poincaré-Nancy 1, UMR 7555, , B.P. 239, 54506 Vandoeuvre, France
[c] Umeå University, Dept. Experimental Physics, S-90187 Umeå, Sweden

The 2D Fermi surface of 1st stage PdAl$_2$Cl$_8$ acceptor-type graphite intercalation compounds (GICs) has been investigated using the Shubnikov-de Haas (SdH) effect. One fundamental frequency is observed, the angular variation of which confirms its strongly 2D nature, as previously found through electrical conductivity measurements. The energy spectrum can be described by the 2D band structure model proposed by Blinowski et al. We obtain the following parameter values: intraplane C-C interaction energy $g_0$ = 2.7 eV, Fermi energy $E_F$ = –1.1 eV and carrier density. $n_{SdH}$ = 1.1×10$^{27}$ m$^{-3}$. Some fewer details are presented on stage 2 and 3 materials.

## INTRODUCTION

The aim of this work was to determine the band structure of 1$^{st}$ stage $PdAl_2Cl_8$ acceptor-type GICs, a material that has attracted our attention over the past years due to many interesting characteristics including: (i) the high anisotropy ratio in the electrical conductivity $s_a/s_c = 1$-$2\times10^6$ at $T= 4.2$ K[1], compared to a value of $10^4$ in pure graphite; (ii) a transformation under pressure from 1$^{st}$ to 2$^{nd}$ stage with the unexpected result that the Raman signal is almost identical for both stages, translating the fact that the total charge transfer per graphene layer remains the same[2]. We also found in this latter study that 2$^{nd}$ and 3$^{rd}$ stage samples also transformed to higher stage. In the present work, we have made use of the Shubnikov-de Haas (SdH) effect and compare the experimental results to those of the 2D band structure model proposed by Blinowski et al.[3].

## EXPERIMENTAL

Highly oriented pyrolytic graphite (HOPG)-based $PdAl_2Cl_8$ GICs were prepared using the vapour method in a two-zone quartz tube[4]. By reacting the materials at a temperature of 300 °C for a period of 3 days, pure stage 1 samples were obtained as confirmed by x-ray techniques. The chemical composition is $C_{22}PdAl_2Cl_{8.5}$. To prepare the 2$^{nd}$ or 3$^{rd}$ stage compounds, we heated (at 320°C, under chlorine) a mixture of graphite and the chloroaluminate with a gravimetric ratio deduced from the stage 1 chemical composition.

The samples (~$5\times1\times0.5$ mm$^3$) were mounted on a rotating sample platform in order to enable angular dependent measurements. The angle *q* between the magnetic field and the c-axis could be determined within 0.2°. Voltage and current leads were attached to the sample by silver paint under a protecting nitrogen atmosphere. The transverse magnetoresistance was measured with a standard four-point dc technique for currents, $I< 0.2$ A, directed in the hexagonal plane (i.e. along the long axis of the sample). High-magnetic fields were produced in the 40 T long-pulse (1 second pulse duration) magnet of the University of Amsterdam. SdH data were obtained in the free decay mode, after ramping the field to its maximum value. Experiments were carried out at $T= 4.2$ K and 1.5 K with the samples immersed in liquid helium in order to ensure stable temperatures.

**RESULTS**

The primary results of our study on the 1$^{st}$ stage material are shown in figure 1 which gives the magnetoresistance, $R(B)-R(B=0)$, measured at $T= 4.2$ K. Pronounced monochromatic SdH oscillations appear for $B> 10$ T. After subtracting the background contribution, estimated by smoothly extrapolating the low-field data to higher fields, we calculated the Fourier transform which is shown in the inset. A fundamental frequency $F_1$ is found at 1190 T. The small peak $F_2$ at 2380 T is its second harmonic. From the period of the SdH oscillation, the extremal cross-section of the Fermi surface $S= \pi(k_F)^2 = (2\pi e/\hbar)F$ and the carrier density $n_{SdH} = 4S/(2\pi)^2 I_c$ can be calculated. We find $n_{SdH} = 1.10\times 10^{27}$ m$^{-3}$.

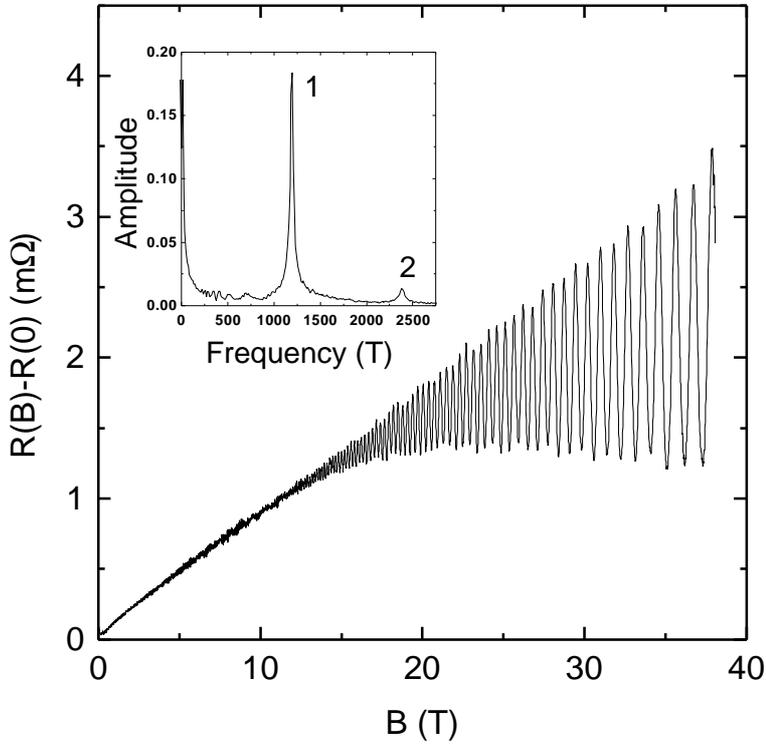

FIGURE 1 Magnetoresistance of the 1$^{st}$ stage PdAl$_2$Cl$_8$ GIC at $T= 4.2$ K. Inset shows the Fourier spectrum of the SdH oscillations.

In order to determine whether the whole Fermi surface has been observed in the SdH experiment $n_{SdH}$ may be compared with the carrier density $n_H= 1/R_H e$, where the Hall constant $R_H= V_H d/IB$ is calculated from the low-field Hall data. Here $V_H$ is the Hall voltage and $d$ is the thickness of the sample. However, we did not obtain reliable values for $R_H$, which we attribute to the non ideal Hall voltage contacts on the sample. The contacts were made by gluing thin copper leads to the side (thickness $d$) of the sample by means of silver paint. Apparently, the thickness of the conducting layer is not identical to the physical thickness of the sample, as we obtained values for $n_H$ about a factor 10 lower than $n_{SdH}$. However, the SdH data of the 1$^{st}$ GICs can be well accounted for in the proposed band structure model (see below). Therefore, we believe that our SdH data disclose the complete Fermi surface.

The angular dependence of the SdH frequencies was measured at 4.2 K. The SdH effect could be observed up to $q$ ~25° and the fundamental frequency $F_1$ followed the relation $F_1(\theta)= F_1(0)/\cos q$, as expected for a 2D cylindrical Fermi surface.

Cyclotron effective masses were determined from the temperature dependence of the amplitude of the SdH oscillations. Since data were taken only at $T = 4.2$ and 1.5 K the accuracy is limited but we find that the effective mass equals $m^*/m_0= 0.25\pm0.05$ ($m_0$ is the free electron mass).

Data were also taken on stage 2 and 3 materials. Non-monochromatic SdH oscillations appear in both cases for $B> 10$ T. The Fourier transform of the data show six and seven frequency components respectively. The angular dependence of the SdH frequencies for the 2$^{nd}$ stage material obeyed the same relationship as for the 1$^{st}$ stage GICs, suggesting a 2D Fermi surface; however, this was no longer the case for the 3$^{rd}$ stage sample.

## ANALYSIS AND DISCUSSION

In this section, we will mainly deal with the results concerning the first stage materials. For strongly anisotropic acceptor-type GICs, Blinowski et al.[3] proposed a two-dimensional band structure model making use of a tight binding method and taking into account interactions between nearest neighbour carbon atoms. The stage 1 GIC is treated as a collection of equivalent independent subsystems consisting of a graphite layer sandwiched between two intercalant layers in which electron transfer from the carbon atoms to the acceptor intercalant results in free holes in the graphene layer. The band structure corresponds to the 2D band structure of graphite[5] with the in-plane interactions between nearest neighbour carbon atoms taken into account only. In the vicinity of the Brillouin zone edge (K-point) the energy dispersion law becomes a linear function of the wave number $k$:

$$E_{c,v}(k) = \pm \frac{3}{2} \gamma_0 b k \tag{1}$$

Here $\gamma_0$ is a parameter which describes the carbon nearest-neighbour interaction energy, as defined by the resonance integral for the overlap of wave functions at neighbouring carbon atoms, and $b$ (= 1.42 Å) is the carbon nearest-neighbour distance. The subscripts $c$ and $v$ label the conduction and valence bands, respectively. The Fermi energy is given by:

$$E_F = -\frac{3}{2} \gamma_0 b \left(\frac{S}{\pi}\right)^{1/2} \tag{2}$$

where $S = \pi k_F^2$ is the extremal Fermi surface cross-section, which in turn can be calculated from the measured frequency $F$ of the SdH oscillation, $S = (2\pi e/\hbar)F$. The cyclotron effective mass is expressed in the Fermi energy as follows:

$$m^* = \frac{\hbar^2}{2\pi} \frac{\partial S}{\partial E} = \frac{4\hbar^2 E_F}{9\gamma_0^2 b^2} \tag{3}$$

The observation of one single fundamental frequency $F_1$= 1190 T in the SdH signal (Fig.1) is consistent with the Blinowski model. With the experimental values of $F_1$ and $m^*$, $\gamma_0$ is evaluated using eqs. (2)-(3) and we obtain $\gamma_0$= 2.7±0.5 eV. This value is in good agreement with the values $\gamma_0 \approx$ 2.8-3.0 eV obtained for other acceptor GIC by reflectivity experiments. For pure graphite, $\gamma_0$= 3.2 eV[6]. A reduction of the $\gamma_0$ value in such a stage 1 acceptor-type GIC is expected, since the increase of carrier density in the graphene layers results in a stronger screening of the atomic potential. The Fermi energy calculated by inserting $\gamma_0$= 2.7 eV and the value for $F_1$ in eq.(2) amounts to –1.1±0.2 eV. Finally, the charge transfer per carbon atom within the Blinowski model is given by $f/l$= 3√3$b^2S/(4\pi^2)$ and amounts to the relatively small value of 0.030e$^+$.

Analysis of the data for the 2$^{nd}$ and 3$^{rd}$ stage products is considerably more difficult, for two different reasons. First, while the X-ray data clearly indicate a three-layer (i.e., Cl-(Pd+Al)-Cl sandwich structure) for both 1$^{st}$ and 3$^{rd}$ stage materials, the 2$^{nd}$ stage material appears to involve a five-layer Cl-Al-Pd-Al-Cl arrangement[7]. Secondly, for the stage 3 compound, with their undoubtedly lower anisotropy, i.e., their more 3D character, the SdH data cannot be analysed within the framework of the 2D Blinowski model attributed to the additional interactions between the graphite subsets across the intercalant layer.

We conclude that the Blinowski model gives a proper description of the band structure of the richest $PdAl_2Cl_8$ GIC. The applicability of a purely 2D model is consistent with the measured $1/\cos q$ dependence of the fundamental frequency, indicating a perfect cylindrical Fermi surface and the high anisotropy ratio of the basal-plane and c-axis conductivity. The value $s_a/s_c \sim 10^6$ appears to be a prerequisite for pure 2D behaviour. Indeed, SdH data on the stage 1 compounds $C_{9.3}AlCl_{3.4}$ and $C_8H_2SO_4$, with much lower anisotropy ratio's of $10^4$ and $10^5$, respectively[8], could not be analysed satisfactorily within the 2D Blinowski model[9].

**Acknowledgements**
This work was part of the research programme of the Dutch "Stichting FOM" (Foundation for Fundamental Research of Matter). The authors thank D.T.N. de Lang for assistance in analysing the data.